# Formation of yttrium oxalate phase filled by carbon clusters on the surface of yttrium oxide films

D.W. Boukhvalov[1,2,*], D.A. Zatsepin[2,3], D.Yu. Biryukov[2], Yu.V. Shchapova[2], N.V. Gavrilov [2,4], and A.F. Zatsepin[2]

[1] College of Science, Institute of Materials Physics and Chemistry, Nanjing Forestry University, Nanjing 210037, P. R. China
[2] Institute of Physics and Technology, Ural Federal University, Ekaterinburg, Russia
[3] Institute of Metal Physics, Ural Branch of Russian Academy of Sciences, Ekaterinburg, Russia
[4] Institute of Electrophysics, Ural Branch of Russian Academy of Science, Ekaterinburg, Russia

**Abstract:** In the current paper, we report the results of surface modification of cubic $Y_2O_3$ films employing carbon-ion implantation. The characterization results demonstrate the formation of a stable yttrium oxalate-based structure with cavities filled with carbon clusters. Theoretical simulations demonstrate that the incorporation of eighteen-atom carbon clusters into the cavities of $Y_2(C_2O_4)_3$ does not lead to valuable changes in the crystal structure of yttrium oxalate. X-ray diffraction and optical measurements demonstrate that the subsurface bulk area of cubic yttrium oxide remains unperturbed. The oxalate "skin" thickness with embedded carbon clusters is estimated to be approximately 10 nm. The prospective employing the method to manage optical properties and increase the biocompatibility of yttria and lanthanide oxides are discussed.

E-mail: danil@njfu.edu.cn

## 1. Introduction

Composites of carbon nanodots (from now on referred to as CNDs) and metal oxides have been intensively studied in recent years due to their promising applications in storage materials [1], sensing [2, 3], solar cells [4], and especially catalysis [5-10]. The combination of wide bandgap in oxides and narrower bandgaps in CNDs having different chemically active centers in both parts of such composites predicts that these materials will be equally important both for photovoltaic and photocatalytic applications [4, 7-10]. At the same time, one of the limitations hindering this important area's onward development is the composites' stability due to the weak adhesion of CNDs to metal substrates [1, 2, 9, 11]. Another approach to improve the future application of these composites is to reduce the degradation of functional properties of oxides in the presence of carbon nanosystems [12]. The biological activity of some CNDs-metal-oxide composites has recently been reported (see, e.g., Ref. [13]), which may also be a potential source of cytotoxicity. As one of the possible solutions to this problem, the incorporation of nano-carbons into the surface of the metal oxide host is proposed. The results reported in Refs. [14-16] indicate a simple dissolution of carbon in oxide host matrices without any formation of CNDs. At the same time, unfortunately, only a limited number of both oxides (usually ZnO [17, 18] and $TiO_2$ [19]) and a limited number of fabrication methods have been considered, which is not sufficient to clarify the actual situation. Thus, employing other incorporation methods of carbon atoms into the other oxide hosts seems to be a reasonable solution at this stage.

Yttrium oxide $Y_2O_3$ is a rare earth sesquioxide with several promising physical properties that allow us to consider this material as a suitable technological host-oxide for antireflection protective coatings, doped laser applications, barrier coating of high-temperature composites, electric gate interfaces, etc. (see, e.g., Refs. [20-24]). The above properties usually include a relatively wide bandgap, high melting point, low thermal expansion, and good phase stability. At the same time, the latter feature belongs only to the cubic bixbyite polymorph of yttrium oxide (*C*-type $Y_2O_3$ or *C*-$Y_2O_3$). In contrast to bixbyite, hexagonal and monoclinic phases can be fabricated only at temperatures above 2325 °C [25] and relatively high pressures [25, 26]. One of the other potential applications of yttrium oxide nanoparticles is bioimaging [23, 27-29]. Many studies of the potential toxicity of yttrium oxide and the use of yttrium oxide nanoparticles as biomarkers have revealed essential suppression of oxidative stress in various living creatures due to the performed injection of yttrium oxide nanoparticles [30-33]. The release of yttrium ions (considered free-radical harvesters) from the surface of $Y_2O_3$ nanoparticles was discussed as a leading source of this effect [34, 35]. Additionally, large-scale application of yttria-based nanoparticles for bioimaging requires an increase in the biocompatibility of these materials [36, 37]. Therefore, modifying the yttrium oxide surface through carbon incorporation may have a substantial practical impact beyond what was discussed in the previous paragraph regarding catalytic and sensing properties.

The *C*-type atomic structure has an *Ia*3 space group with 16 formula units in the elementary cell [25, 38]. Such structural feature is well-suitable for substitution doping with rare-earth ions of lattice-sites yttrium atoms. However, it is not compatible with other ions that are not rare-earth [22, 39]. Since the *C*-type $Y_2O_3$ unit cell contains 48 oxygen and 32 yttrium ions located in the center of an approximate cube, interstitial doping is preferred for other ions, and carbon atoms are not an exception. Therefore, cubic $Y_2O_3$ can be considered as a natural candidate for testing the possibility of CNDs formation in a metal-oxide host-matrix.

Ion implantation is a powerful technique for surface modification [40,41]. Typically, this method can be employed to fabricate point defects (single impurities and clusters) in metals [42] and semiconductors [43,44]. Among the oxide host-materials ZnO, $TiO_2$ and $SiO_2$ were usually chosen as the ordinary substrates for implantation. Surface implantation of heavier metals oxides such as $Gd_2O_3$ has been the subject of a small number of studies in which low concentrations of impurities are embedded into the hosts [45,46]. Some scientific papers also report the formation of a secondary phase in semiconductor host-substrate [47].

In this paper, we report the results of synthesis of cubic yttrium oxide films employing magnetron sputtering and onward incorporation of significant amount of carbon impurities into the surface by means of pulsed ion implantation. The modified films were systematically studied by Scanning Electron Microscopy (SEM), X-Ray Diffraction (XRD), X-Ray Photoelectron and Raman spectroscopy, followed by measurements of optical properties. Experimental results were used for theoretical simulation of carbon nano-clusters embedded into yttrium oxalate phase, formed on the surface of yttrium oxide after implantation of relatively significant amount of carbon ions there.

## 2. Experimental details and theoretical method

An yttrium oxide film was deposited on a quartz glass substrate employing constant pulse mode (50 kHz, 10 μs) with the help of reactive magnetron sputtering technology. Immediately before the deposition process, the quartz glass substrate was cleaned in acetone solvent using an ultrasonic bath for 20 min. Then, air drying of the blank substrate was performed. An yttrium source was obtained by cold pressing of yttrium metal powder of 99.99 % purity at 30 MPa pressure, yielding the source geometry 40 mm in diameter and 2 mm thick. The magnetron, the sputtered source, and the blank substrate were loaded into a vacuum chamber, which then was pumped out to $6.6\times10^{-3}$ Pa pressure using a turbomolecular pump. The deposition of yttrium was carried out at 30 W of magnetron power for 8 h in a combined argon-oxygen atmosphere with at least 30% volume concentration of oxygen. The substrate temperature during the deposition process was limited to 400 ± 25°C. Then, the final sample was cooled down to room temperature in a vacuum chamber. The final thickness of magnetron sputtered $Y_2O_3$ blank (non-implanted) film was ~700 nm, which was determined by the ball abrasion method employing the Calotest system (CSM Instruments SA, Peseux, Switzerland).

Carbon pulsed implantation of synthesized and stored in ambient conditions $Y_2O_3$ thick film had been carried out as follows. A beam of C-ions with the energy of 30 keV was generated in the repetitively pulsed mode with a pulse duration of 0.4 μs and 6.25 Hz repetition rate. The vacuum in the implanter chamber was not worse than 0.01 Pa and was kept employing a turbomolecular pump. The ion fluence was $10^{17}$ cm$^{-2}$, which is an order larger than typically used for surface modification [45, 47, 48]. The exposure of the sample under a pulsed ion beam was at least 370 seconds. After carbon implantation, the sample was kept in a vacuum chamber for 12 minutes before being placed in a humidity-protected transfer vessel to transfer it to other experimental techniques.

The implanted films deposited on the substrates were visualized using a Carl Zeiss Evo LS10 scanning electron microscope. Structural-phase analysis of the $Y_2O_3$ sample under study was performed using an XPertPro MPD diffractometer employing Cu $K\alpha$ = 1.54 Å radiation (see XRD pattern shown in Fig. 1).

Since yttrium oxide films are widely used in different optical applications (see Introduction section), it is logical to apply surface-sensitive methods to inspect the chemical composition and grade of the surface of our samples. For this reason, we employed X-ray Photoelectron Spectroscopy (XPS) using survey and core-level analysis with the help of a ThermoScientific *K-alpha Plus* XPS spectrometer. This spectrometer has a monochromatic microfocused Al $K\alpha$ X-ray source and has 0.05 at% element sensitivity. Operating pressure in an analytic chamber during measurements was not worse than $1.3 \times 10^{-6}$ Pa. Dual-channel automatic charge compensator (GB Patent 2411763) was applied to exclude the charging of our sample under XPS analysis because of the loss of photoelectrons. Pre-run-up procedures performed included standard degassing of the sample and analyzer binding energy scale inspection and re-calibration (if needed) employing sputter-cleaned Au ($4f_{7/2}$ band), Ag ($3d_{5/2}$ band), and Cu ($2p_{3/2}$ band) inbuilt XPS Reference Standards according to ISO 16.243 XPS International Standard and XPS ASTM E2108-00 Standard. We used a ThermoScientific XPS spectrometer inbuilt electronic database and a well-known approved NIST XPS Standard Reference Database (USA).

Raman spectra were obtained in a backscattering geometry with the help of Horiba LabRam HR800 Evolution spectrometer equipped with an Olympus BX-FM confocal microscope with 100×/NA = 0.9 lens and 514.5 nm line of argon-ion laser ("Geoanalyst" Shared Equipment Center, Institute of Geology and Geochemistry, Ural Branch of Russian Academy of Sciences). The registration was carried out via Czerny-Turner monochromator with 600 grooves/mm diffraction grating employing a multichannel electrically cooled CCD detector. The measurements were carried out with a confocal aperture of 50 μm. Spatial axial resolution was

~1.7 μm. To analyze the structure of the submicron layers of the film, the 3D arrays of Raman spectra were collected by the z-depth profiling with 0.3 μm step [49].

Optical transmission and absorption spectra were recorded with the help of Perkin Elmer Lambda 35 spectrophotometer using an integrating sphere in the 190 – 1100 nm spectral range. The luminescence and excitation spectra of photoluminescence were recorded on a pulsed luminescence spectrometer Perkin Elmer LS 55 in the phosphorescence mode. The size of the registration time window was 12 ms, and the delay was 0.04 ms. Luminescence spectra were recorded in the 300 – 900 nm spectral range with an optical filter of 290 nm. Photoluminescence excitation spectra were recorded in the 200 – 350 nm spectral range with filters of 350 and 390 nm.

Theoretical modeling was performed using the QUANTUM-ESPRESSO pseudopotential code [50], employing the generalized gradient approximation [51] for exchange-correlation potential in a spin-polarized mode. A full optimization of atomic positions and lattice parameters was carried out. During optimization, the electronic ground state was consistently found using ultra-soft pseudopotentials [52].

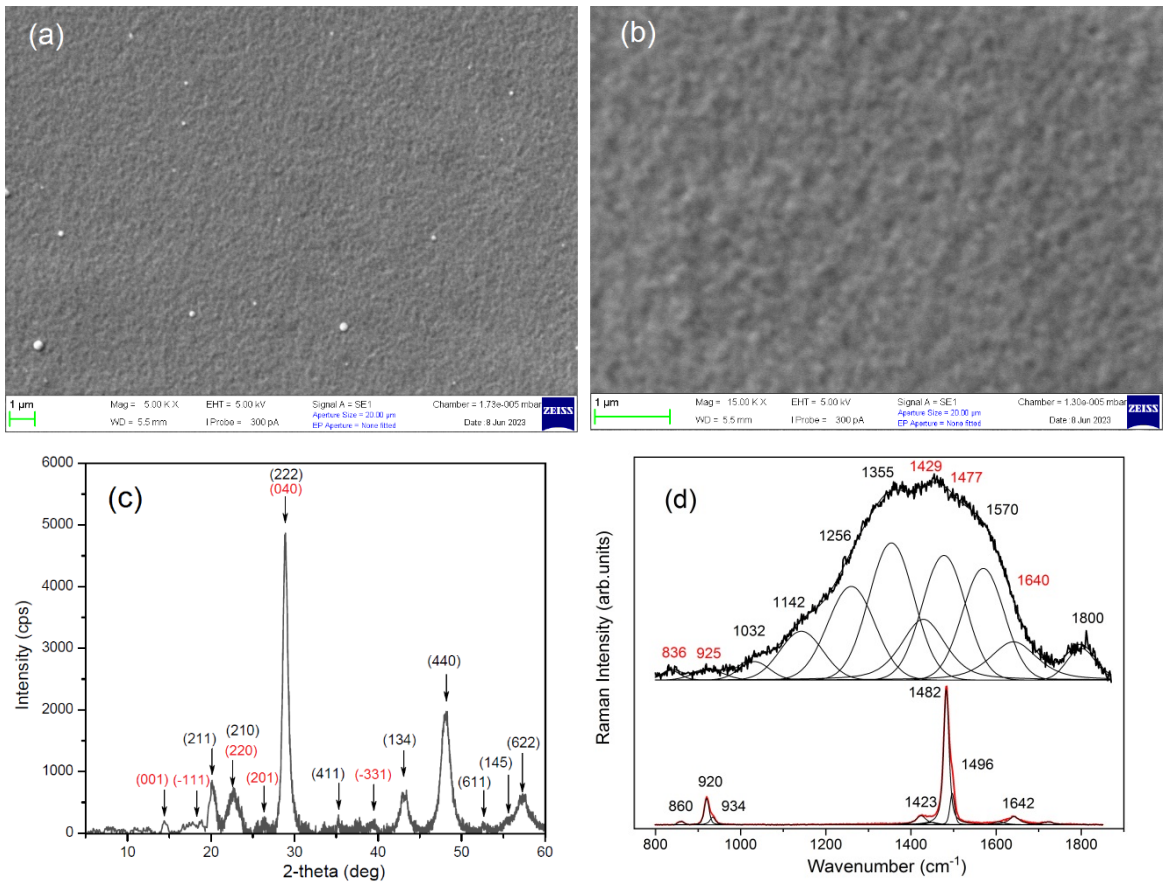

**Figure 1.** SEM images of the surface (a,b), XRD patterns (c) and Raman spectra (800–1850 cm$^{-1}$) of $Y_2O_3$:C (black) and $Y_2(C_2O_4)_3 \cdot nH_2O$ (red) together with their deconvolution using Voigt functions (d). Red numbers on panels (c,d) indicate features linked with $Y_2(C_2O_4)_3$ phase.

## 3. Results and discussion

### 3.1. Samples characterization

The first stage of our study is to inspect the surface homogeneity employing Scanning Tunneling Microscopy (SEM). The SEM images shown in Figure 1(a-b) persuasively demonstrate the homogeneity of the surface under study. One can conclude from the XRD patterns presented in Fig. 1c that the cubic $Y_2O_3$ structure is dominating in the inspected sample. The (222) XRD reflex is characterized by the highest intensity with interplanar peak spacing, corresponding to cubic lattice parameter $a = 10.70$ Å [53]. We have to note that all XRD peaks are characterized by relatively significant width (FWHM = 0.75 – 1.1 deg). Apparently, this can be explained by the noticeable disordering of the initial $Y_2O_3$ cubic lattice after carbon implantation. Also, two weak peaks located in the 14.6 and 26.4 degrees region are not associated with the $Y_2O_3$ cubic lattice and will be interpreted onward.

Raman spectrum of $Y_2O_3$:C sample in the first-order of carbon vibrations area demonstrates a complex superposition of highly broadened bands 836, 925, 1032, 1142, 1256, 1355, 1429, 1477, 1570, 1640, ~1800 см$^{-1}$ (Fig. 3d). This spectrum can be attributed to typical carbon vibration bands of G, $D_1$, $D_2$, $D_3$, D″ and D** (Tab.1). The nature of the bands mentioned was described in numerous publications (see, e.g. Refs. [54-57]). The G-band corresponds to Raman-allowed phonon mode at the center of the Brillouin zone with $E_{2g}$ symmetry and is governed by a single resonance process. It arises from the stretching vibration of *sp²*-bonded pairs and indicates a graphitization of material. The $D_1$ and $D_2$ bands are associated with sp³ defects (disorder), activating the double resonance scattering of single phonons outside the center of the Brillouin zone due to the possibility of electron-hole relaxations. The $D_3$ band and the wide bands near 1200 cm$^{-1}$ are specific and belong to defective carbons, so they were assigned to the presence of amorphous carbon [58,59].

However, it should be noted that the spectra of the $Y_2O_3$:C sample are different from those for mechanically activated and nano-graphites [60,61]. In particular, this difference consists in the high relative intensity of 1429 and 1477 cm$^{-1}$ bands and the presence of weak bands in the region of less than 1000 cm$^{-1}$. The C–O, C–OH, C=O, and $H_2O$ bonds inherent vibrations of $Y_2(C_2O_4)_3·nH_2O$ structure and may appear in these areas [62,63] (see Fig. 3d and

Tab.1). Accordingly, the presence of $Y_2(C_2O_4)_3 \cdot nH_2O$ phase in the surface layers of $Y_2O_3$:C sample can be hypothesized.

**Table 1.** Peak positions and type of vibrations in Raman spectra (800-1850 cm$^{-1}$) of $Y_2O_3$:C, $Y_2(C_2O_4)_3 \cdot nH_2O$ and mechanically activated graphite.

| Peak position, cm$^{-1}$ | | | |
|---|---|---|---|
| $Y_2O_3$:C [a] | $Y_2(C_2O_4)_3$ [b] | [Gd$^{3+}$]/[oxalate] [b] | Mechanically activated graphite [c] |
| 836 | 860 | 855 ($\delta$(C-C)+ $\nu$(C-O)) | - |
| 925 | 920 | - | - |
|  | 934 | - |  |
| 1032 |  | - | - |
| 1142 | - | - | 1110-1140 (D'') |
| 1256 | - | - | 1260 (D**) |
| 1355 | - | - | 1353 (D$_1$) |
| 1429 | 1423 | - | - |
| 1477 | 1482 | - | 1480 (D$_3$) |
|  | 1496 | 1490 ($\delta$(C-OH)+ $\nu$(C-O)) |  |
| 1570 | - | - | 1580 (G) |
| 1640 | 1642 | 1628 ($\nu$(H$_2$O)) | 1620 (D$_2$) |
| ~1800 | 1725 | 1738 ($\nu$(C=O)) | - |

Note: [a] this work, [b] Ref. [63], [c] Ref. [60].

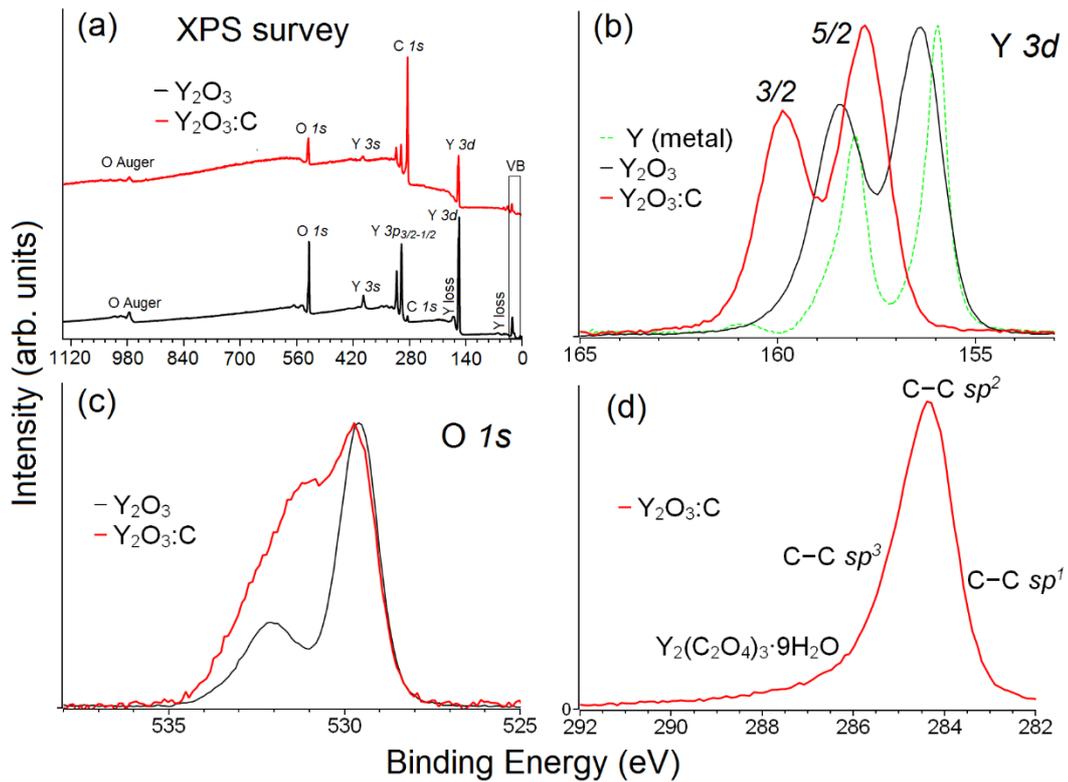

**Figure 2.** XPS survey (a), Y 3*d* (b), O 1*s* (b), and C 1*s* (d) core-level spectra of $Y_2O_3$:C in comparance with appropriate XPS external standards.

XPS survey and core-level spectra are shown in Fig. 2. The survey spectra in Fig. 2a demonstrate the absence of contaminations in the range of XPS spectrometer sensitivity (recall that it has 0.05 at. % element sensitivity limits). The spectrum of yttrium 3*d* orbitals of the sample under study (Fig. 2b) differs significantly from the usual spectra of cubic yttria. Based on the comparison mentioned, we can propose significant changes in the atomic structure of $Y_2O_3$:C in the surface and surface-adjacent areas. The simultaneous increase in the binding energies corresponding to Y 3*d* orbitals can be attributed to the formation of yttrium oxalate $Y_2(C_2O_4)_3$ phase (see Fig. 3a), where Y $3d_{3/2}$ and Y $3d_{5/2}$ peak positions are 161.6 eV and 159.5 eV, respectively [64]. The helicoidal channels of yttrium oxalate (see Fig. 3) and other lanthanides oxalates can be filled with water molecules to form oxalate nonahydrates ($Y_2(C_2O_4)_3 \cdot 9H_2O$). [65]. Incorporation of water into the cavities of yttrium oxalate is stable up to 45°C [66]. Such kind of hydration leads to the peaks-shift to 160.5 eV and 158.1 eV [67]. These binding energy values for the sub-bands mentioned are in good agreement with the actually measured Y 3*d* spectrum. The nature of Y 3*d* peaks-shift after filling the pores will be discussed in the next section of the current paper. To prove the conclusion about oxalate phase formation, we inspected XRD data shown in Fig. 1c. We found that peak patterns at 14.6 and 26.4 degrees and several other less distinct features are very similar to that for $Y_2(C_2O_4)_3$ [68]. This conclusion is consistent with the above assumption about the participation of the Y$_2$(C$_2$O$_4$)$_3$ phase in shaping the Raman spectrum in the region of first-order carbon vibrations. Thus, we can claim with a high degree of probability that at least some part of implanted carbon atoms incorporates into the surface of yttrium oxide as Y$_2$(C$_2$O$_4$)$_3$ surface phase. Note that Y 3$d$ spectra demonstrate the total absence of Y$_2$O$_3$ signatures. Since the depth resolution in the XPS method is 7 nm, we can claim that the thickness of oxalate-like skin is more than 7 nm.

XPS O 1$s$ spectrum of Y$_2$O$_3$:C shown in Fig. 2c also demonstrates a significant deviation from that for initial (unimplanted) cubic yttrium oxide toward spectra of Y$_2$(C$_2$O$_4$)$_3$·9H$_2$O where the corresponding peak is usually located at 531.4 eV [67]. Note that in non-hydrated yttrium oxalate, the O 1$s$ peak is located at 533.5 eV [64], which is in less agreement with measured spectra. Thus, Y 3$d$ and O 1$s$ spectra conform to the proposed model of the formation of an oxalate-like phase with moisture-filed pores in the Y$_2$O$_3$:C sample.

A permeation of moisture into the cavities of yttrium oxalate leads to the shift of C 1$s$ core-level from 290.1 eV to 288.5 eV with significant smearing of this peak, which makes it almost undistinguished from the background [64,67]. Thus, the long tail in the XPS C $1s$ spectrum shown in Fig. 2d can be linked to the spectral contribution of carbon atoms from Y$_2$(C$_2$O$_4$)$_3$·9H$_2$O-like structure. The band with a distinct maximum at about 284.3 eV is related to some carbon-based structures. The band mentioned is mainly associated with carbon atoms with $sp^2$ hybridization. Visible contributions from carbon atoms with $sp^3$ and $sp^1$ hybridizations were also detected and established. Alternatively, several interpretations of this peak can be proposed. The first is the association of this peak with organic contaminants.

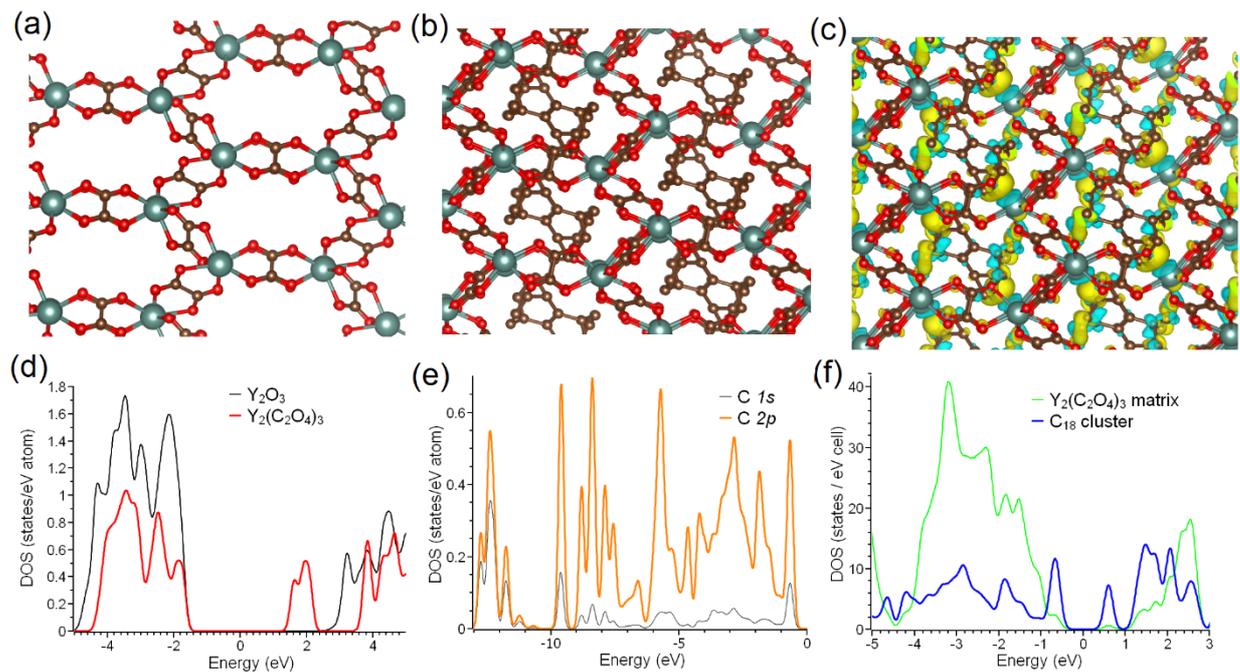

**Figure 3.** Optimized atomic structure of $Y_2(C_2O_4)_3$ before (a) and after (b) fabrication of carbon clusters ($C_{18}$) in cavities. Yttrium atoms are shown as large grey-blue spheres, and oxygen and carbon as small red and brown spheres. The calculated charge transfers between carbon clusters and $Y_2(C_2O_4)_3$ matrix (c). The yellow "clouds" correspond to the increased densities of charge and cyan to the decreased one. Total densities of states of yttrium oxide and oxalate (d). Partial densities of states carbon atom with *sp²* hybridization from $C_{18}$ cluster incorporated into $Y_2(C_2O_4)_3$ matrix (e). The total densities of states of carbon clusters and $Y_2(C_2O_4)_3$ matrix after incorporation are shown in panel (f). Fermi energy is set as zero and does not coincide with zero energy in XPS measurements shown in Figs. 2 and 3.

We can rule out this interpretation since such characteristic features in the measured C 1*s* core-level spectrum are absent over 289 eV due to low-energy soft-sputtering performed directly before XPS measurements according to XPS ASTM E2108-00 Standard. Another possible interpretation is the formation of graphite-like structures (for example, nano-graphenes). These graphite-like clusters have distinct Raman signatures, even for small and distorted structures [55]. At the same time, Raman spectra shown in Figure 1d demonstrate the absence of distinct *D* and *G* peaks. Thus, the formation of some carbon structures with graphene-like layers can be ruled out. The following interpretation, which can be well-supported by Raman spectra, is the formation of some amorphous carbon-based structures. The typical feature of amorphous carbon is the visible extension of the valence band edge upon the Fermi level [69]. XPS valence band spectra shown in Fig. 4 demonstrate an insignificant shift of the valence band edge toward the Fermi level. Thus, the formation of amorphous carbon can also be ruled out. In addition, the sample was washed in acid before XPS measurement. This treatment provides the removal of all discussed above surface contaminants. Since experimentally obtained spectra are close to that for yttrium oxalate with pores filled by the water, it is reasonable to propose the formation of some carbon clusters inside the pores of $Y_2(C_2O_4)_3$, as shown in Fig. 3b. Note that incorporation of carbon pairs into the voids of Y-based porous systems was already observed earlier for $Y_{10}I_{13}C_2$ [70].

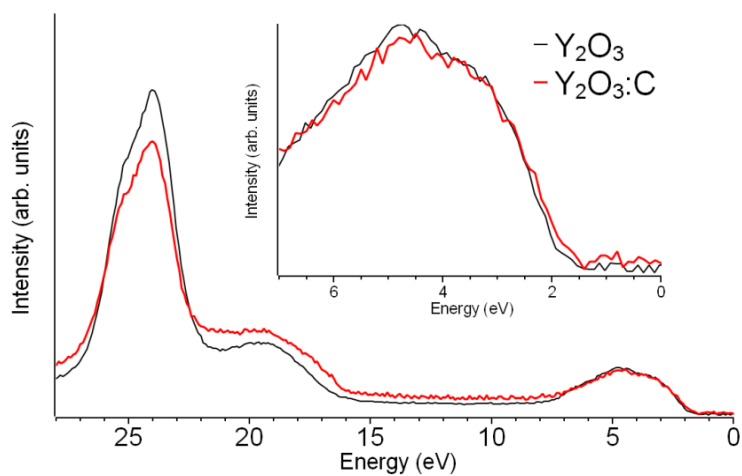

**Figure 4.** XPS spectra of the valence band and valence band edges of $Y_2O_3$ and $Y_2O_3$:C.

### 3.2. Theoretical modeling

The set of first-principles calculations was carried out to support the model of $Y_2O_3$:C phase fabrication based on XRD, XPS, and Raman data combined analysis. In the first stage, we simulated how the formation of surface oxalate skin can affect the valence band structure. Calculations performed, and their data shown in Fig. 4d demonstrate the coincidence of valence bands edge of yttrium oxide and oxalate. At the next stage, we simulated carbon nanoclusters with some carbon atoms having mostly $sp^2$ hybridization. Since helicoidal pores of oxalate nonahydrates can incorporate an additional twelve water molecules [65], we started our simulation with clusters consisting of ten atoms. We proposed several models with small fullerene-like carbon clusters of size smaller than cavities in $Y_2(C_2O_4)_3$ and containing several carbon atoms. After that, a full optimization of lattice parameters and atomic positions of yttrium oxalate with incorporated carbon clusters was carried out. Several model clusters collapsed or formed covalent bonds with the surrounding host matrix. One of the clusters shown in Fig. 3b survived during the optimization of atomic structure. Note that incorporating $C_{18}$ clusters into the cavities does not provide visible changes in the overall crystal structure of the oxalate matrix. The changes in lattice parameters are also less than 10%, which is of the same order of magnitudes as in commercial $Y_2(C_2O_4)_3 \cdot 9H_2O$ [71]. Thus, the results of simulations are in qualitative agreement with XRD data.

To inspect carbon atom hybridization, partial densities of states have been calculated. Results of the calculations shown in Fig. 3e demonstrate the $sp^2$ hybridization pattern with a significant number of states near the edge of the valence band, which is typical for carbon atoms. Thus, the simulated structure can be used as a model of yttrium oxalate with an embedded carbon system with a visible number of $sp^2$ hybridized carbon atoms. Calculated total densities of states of $C_{18}$ clusters and $Y_2(C_2O_4)_3$ matrix after implantation are shown in Fig. 3f. The calculations demonstrate insignificant changes in the overall electronic structure of yttrium oxalate. In the electronic structure of $C_{18}$ clusters, one can see distinct peaks on the edges of the energy gap. Note that this distinct peak in the valence band is located precisely on the edge of the valence band of the oxalate matrix. Thus, the tiny shift of the valence band edge shown in Fig. 4 can be interpreted as a contribution from carbon clusters incorporated as a contribution from carbon nanoclusters embedded into the oxalate matrix. To explain the deviation of XRD patterns of $Y_2O_3$:C from $Y_2(C_2O_4)_3$-like to $Y_2(C_2O_4)_3 \cdot 9H_2O$-like, we calculated the transformations of the charge density distribution after incorporating the $C_{18}$ cluster.

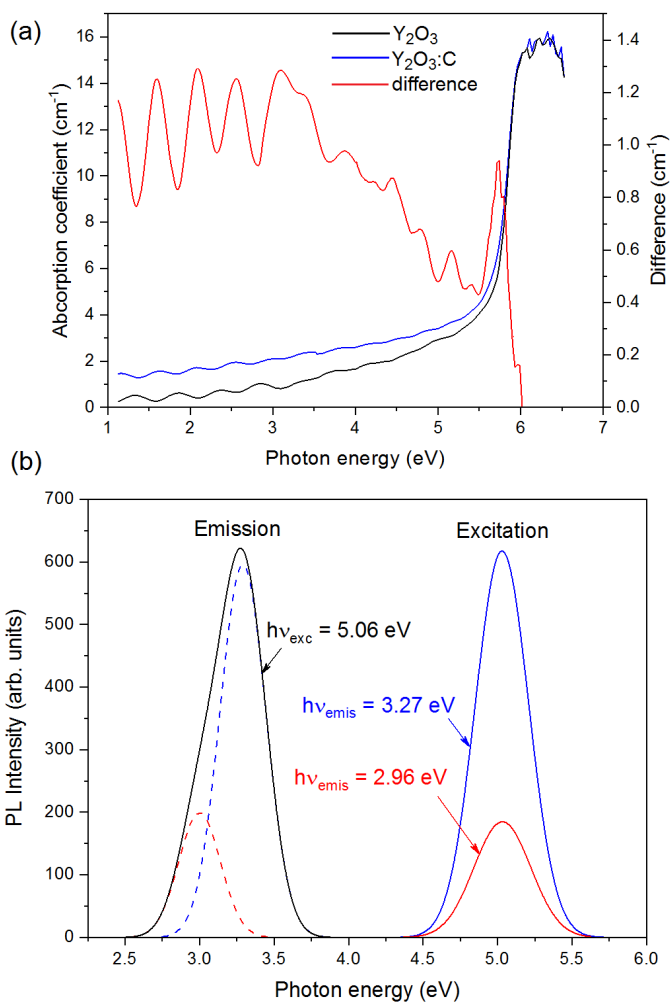

**Figure 5.** Optical absorption spectra of $Y_2O_3$ and $Y_2O_3$:C and the difference between measured absorption coefficients (a). Spectra of photoluminescence ($h\nu_{exc}$ = 5.06 eV) and photoluminescence excitation ($h\nu_{emis}$ = 3.27 eV and 2.96 eV) of $Y_2O_3$:C sample (b).

Calculations data shown in Fig. 3c demonstrate visible charge densities redistribution in both carbon cluster and oxalate matrix. This charge redistribution explains the origin of transformations in the electronic structure of the oxalate matrix after incorporation of carbon cluster (see Fig. 3d,f) and deviation of XPS spectra from measured for pristine $Y_2(C_2O_4)_3$. In summary of this theoretical section, we want to note that the proposed structure is the only one from a number of stable $sp^2$-carbon clusters that can be fitted into the cavities of yttrium oxalate. However, we propose that the main physical properties of other $sp^2$-carbon clusters will be similar due to the similarity of size and hybridization.

### 3.3. Optical measurements

The last stage of our study is the inspection of the optical properties of the $Y_2O_3$:C sample. Absorption spectra of initial (unimplanted) and carbon-doped yttrium oxides demonstrate a tiny decrease of the bandgap from 5.76 eV (this is very close to 5.78 eV reported for cubic $Y_2O_3$ films prepared by a different method [72]) in undoped samples to 5.74 eV after carbon incorporation. The spectral shift mentioned can be associated with the shift of the valence band edge shown in Fig. 4. Since the calculation results presented in the previous section demonstrate a significant decrease of the bandgap in yttrium oxalate with embedded carbon clusters, we can claim that formation of this structure occurs only in the thin surface part of $Y_2O_3$ sample. Therefore, the contribution of this structure to overall absorption properties is minuscule.

The size of the transparency gap usually depends on the length of the tailing zone in materials of the same composition, and it is determined by the degree of structural disorder [47,73]. Both doped and non-doped samples demonstrate the presence of long tails in absorption spectra. In the case of $Y_2O_3$ films, this tail is associated with unavoidable defects of surface and subsurface areas. To inspect the influence of carbon doping on surface-related optical properties, we calculated the difference between measured absorption spectra of carbon-doped and non-doped yttrium oxides. The calculation results shown in Figure 5a demonstrate the visible difference in absorption spectra on the whole length of the tailing zone. This difference is linked to the appearance of the surface oxalate phase with embedded carbon clusters discussed in previous sections (see discussions above).

Using interference extremes, it is possible to estimate the thickness of the $Y_2O_3$ film using the expression:

$$d = \frac{\lambda_1 \lambda_2}{2(\lambda_2 n_1 - \lambda_1 n_2)}$$

where $n_1$ and $n_2$ are the refractive indices at wavelengths $\lambda_1$ and $\lambda_2$, which correspond to the maxima (minima) of neighboring interference phases. The average value of the film thickness, determined by extremes in the range of 1 eV – 3.5 eV, was 712 ± 28 nm. The analytically obtained value of the film thickness with an error accuracy coincides with the declared at the synthesis stage (see section 2). Based on the magnitudes of absorption coefficients in bulk and surface-related parts of the spectra in Fig. 5a, we can estimate the thickness of the carbon-rich surface skin as about a few dozen nanometers.

Figure 5b shows measured photoluminescence and excitation spectra of $Y_2O_3$:C films. An asymmetrical photoluminescence band with a maximum of 3.26 eV was recorded using an excitation energy of 5.06 eV. This band can be fitted by Gaussian components (see dashed lines in Fig. 5b) in such a way that two components can be distinguished. The first is 3.0 eV (full width at half maximum, further FWHM = 0.306 eV), and the second is 3.3 eV (FWHM = 0.338 eV). The excitation spectra of the PL were recorded at registration energies close to the maxima of the corresponding Gaussian components: 2.96 eV (red maximum) and 3.26 eV (blue maximum). The resulting excitation band in both cases is described by a Gaussian function with a maximum of 5.03 eV. At the same time, the half-width of the band in the first case is FWHM = 0.449 eV, and in the second, FWHM = 0.421 eV, respectively, which may indicate the different nature of detected photoluminescence centers.

For different morphologies of cubic $Y_2O_3$ (bulk crystal, transparent ceramics), luminescence at 3.4–3.5 eV is associated with the luminescence of self-trapped excitons (STE) [74,75]. We propose that the dominant emission band located at 3.3 eV is also associated with the radiative recombination of STE in the $Y_2O_3$ film, and the spectral position of this band is shifted to the region of low energies compared to bulk cases. At the same time, as the literature data show, photons with energy in the region of 5 eV in $Y_2O_3$ crystals with a cubic lattice can also excite photoluminescence caused by radiative recombination of a bound exciton (BE) localized in the center of F-type anionic vacancies [72,76]. Thus, we can conclude that the emission bands located at 3.0 eV and 3.3 eV can be attributed to BE and STE, respectively.

## 4. Conclusions

In the current paper, we reported the results of carbon-ion implantation into the surface of cubic yttrium oxide using $10^{17}$ cm$^{-2}$ ion fluence and onward characterization of the synthesized material. XRD, absorption and photoluminescence measurements demonstrate the stability of cubic yttrium oxide structure with some traces of yttrium oxalate-like phase. XPS measurements demonstrate the formation of $Y_2(C_2O_4)_3 \cdot 9H_2O$-like surface "skin" with cavities filled by some carbon structures with prevailing *sp$^2$* hybridization. No spectral signatures of cubic $Y_2O_3$ have been detected in the XPS spectra of the sample under study. Measured Raman spectra confirm the possibility of oxalate-like phase formation in near-surface layers and indicate a substantially disordered structure of near-surface carbon. Theoretical simulations demonstrate the possibility of incorporating $C_{18}$ clusters into the cavities of yttrium oxalate without visible distortions of crystal structure. The calculated electronic structure of $C_{18}$ clusters demonstrates carbon atoms with *sp²* hybridization. Since the outer layers survived after washing the sample in acid, we can propose chemical stability of the fabricated carbon-rich surface "skin." Note that besides yttrium, all lanthanides also form oxalates with the same pattern of crystal structures. Thus proposed method can be used for suppression of the lanthanides ion leakage from oxide surface, which is essential for increasing the safety of lanthanides-oxides-based devises. Overall, the formation of secondary phase on the surface by means of high-fluence ion implantation is the novel method for the fabrication of few nanometer surface phases on various substrates.

Thus, ion implantation can be used to fabricate carbon-rich skin on the surface of these optically active materials, which should increase the biocompatibility of the materials fabricated from discussed oxides, which is crucial for the material used in bioimaging. Note that yttrium oxalate nonahydrate is a precursor for the synthesis of superconductive $YBa_2Cu_3O_{7-x}$ (YBCO) [77-80]. Thus, $Y_2(C_2O_4)_3 \cdot 9H_2O$-like surface "skin" can also be discussed as a possible precursor for growing some YBCO-based nanosized layers on $Y_2O_3$ substrates.

Measured absorption spectra also demonstrate some changes in optical properties in the surface area. Since the estimated thickness of oxalate-like surface "skin" is only ten nanometers and the thickness of the film is about 700 nm, the contribution of the carbon-modified part of the system is negligible. Decreasing the oxide film's thickness or the nanoparticles' size will increase the contribution from the carbon-rich surface "skin." Since the bandgap of pure and carbon-filled yttrium oxalate is narrower than the bandgap of yttrium oxide, the absorption and emission of visible light can be realized. In general, the obtained data on optical and electronic properties indicate new possibilities for the fabrication general-purpose functional materials for photonic and combined optoelectronic devices, elements of information systems, as well as biomarkers and biosensors.

**Acknowledgments**
The study was supported by the Ministry of Science and Higher Education of the Russian Federation (Ural Federal University Program of Development within the Priority-2030 Program, project 4.38).